# Radiative rates of transitions from the $2s2p^3\ ^5S°_2$ level of neutral carbon

### K Haris[1] and A Kramida

National Institute of Standards and Technology, Gaithersburg, MD 20899, USA

E-mail: haris.kunari@nist.gov

**Abstract**: The measured radiative rates of the $2s^22p^2\ ^3P_{1,2} - 2s2p^3\ ^5S°_2$ intercombination transitions in neutral carbon reported in the literature are critically evaluated by comparing them with theoretical and semi-empirical results. The experimental and theoretical values are compared for the carbon isoelectronic sequence from neutral carbon to nine-times ionized phosphorous. We find strong support for the currently recommended theoretical data on C I and conclude that the published measurements for this transition in neutral carbon cannot be trusted. The reasons for the discrepancies are not clear, and new experiments are needed.

Keywords: neutral carbon, atomic data, intercombination lines, transition rates, radiative lifetimes, isoelectronic comparison

## 1. Introduction

Neutral carbon has a ground $2s^22p^2$ configuration with five levels in it. The level scheme is governed by *LS* coupling, so that $^3P_0$ is the ground level, while $^3P_1$, $^3P_2$, $^1D_2$, and $^1S_0$ are at 16.4 cm$^{-1}$, 43.4 cm$^{-1}$, 10 192 cm$^{-1}$, and 21 648 cm$^{-1}$, respectively. All these levels are almost pure in *LS* coupling. The next excited level is the one with the largest spin quantum number, $^5S°_2$ of the $2s2p^3$ configuration, followed by triplet and singlet levels of $2s^22p3s$, $2s2p^3$, and other configurations involving excitation of electrons from the 2s and 2p shells. The scheme of the lowest energy levels and relevant transitions is shown in figure 1. Because of almost pure quintet character of the $2s2p^3\ ^5S°_2$ level, its radiative decay to lower levels (of $2s^22p^2$), spin-forbidden in *LS* coupling, is very weak, making this level metastable. Spin-changing electric-dipole (E1) transitions, often called intercombination transitions, are usually weak in light atoms. Nevertheless, the $2s^22p^2\ ^3P_{1,2} - 2s2p^3\ ^5S°_2$ transitions in C I (see figure 1) were observed, first by Shenstone in 1947 [1] and later by Johansson [2] under special laboratory conditions. Shenstone's measurements of the intercombination lines, 2964.846 Å and 2967.224 Å (hereafter, lines A and B), established the energy of $^5S°_2$ as 33735.2 cm$^{-1}$. His finding led to identification of these lines in the solar atmosphere [3, 4].

Relaxation of the selection rule requiring preservation of spin in E1 transitions is due to presence of intra- and/or inter-configuration interactions. Because of these interactions, the $2s2p^3\ ^5S°_2$ level is weakly mixed with levels of the same total angular momentum $J = 2$ of other odd-parity terms: $^3P°$ of $2s^22pns$, $^3P°$ and $^{1,3}D°$ of $2s2p^3$ and $2s^22pnd$, and $^5S°$ of $2s2p^2(^4P)np$ and $2p^3ns$ ($n ≥ 3$). In principle, this level can decay to all levels of the ground configuration $2s^22p^2$. The dominant channels defining the radiative lifetime $τ$ of the $2s2p^3\ ^5S°_2$ level are E1 transitions to $^3P_{1,2}$, depicted by arrows in figure 1, while the E1 transition to $^1D_2$, as well as possible magnetic-quadrupole transitions to all levels of $2s^22p^2$ are weaker by several orders of magnitude [5, 6].

---

[1] Guest Researcher

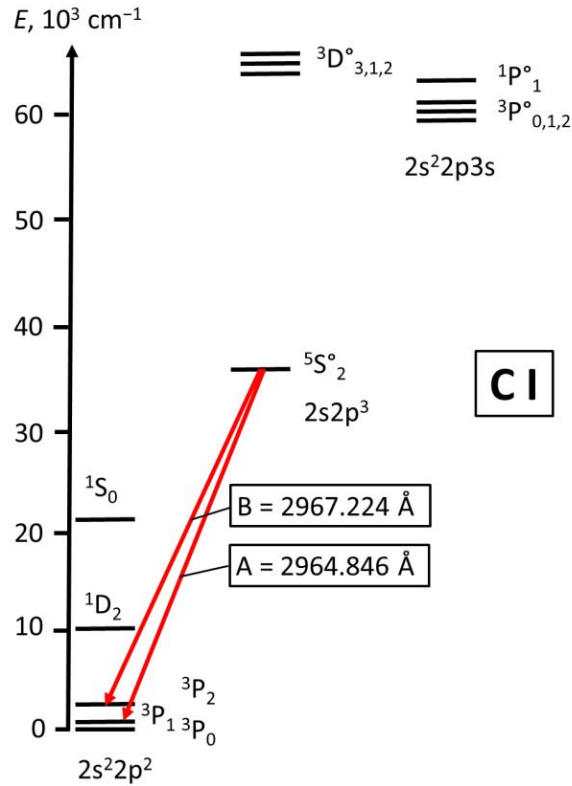

**Figure 1** (*Color online*). The lowest energy levels of neutral carbon (not to scale). The intercombination transitions discussed in this paper are shown by (red) arrows.

Intercombination lines play an important role in a variety of diagnostic applications (e.g., determination of temperature, optical depth, electron density and atomic abundances) for laboratory and astrophysical plasmas. One of the numerous applications of C I transition rates is in determination of atomic number density in carbon evaporation and deposition, which uses ultraviolet (UV) and vacuum ultraviolet (VUV) absorption spectroscopy [7–9]. In general, for many applications a complete set of critically evaluated data on transition parameters, such as transition rate (*A*-value), oscillator strength (*f*-value), and line strength *S* are also required. To a great extent, this requirement is now fulfilled for C I [10–12]. In construction of recommended data sets, experimental data on radiative transition probabilities are commonly used as benchmark in evaluation of theoretical data. However, some exceptions happen when experiments are misinterpreted or contain large unaccounted errors. We found such an exception in the published data on radiative rates of the $2s^22p^2\ ^3P_{1,2} - 2s2p^3\ ^5S°_2$ transitions in neutral carbon. For these transitions, the published measurements by Goly et al. [13] and by Ito et al. [14] grossly disagree with each other and with several consistent theoretical calculations (see [4–5] and references therein). There are also some discrepant theoretical results [15]. To analyze the discrepant data involving the $2s2p^3\ ^5S°_2$ level (either radiative rates or lifetime), it is helpful to compare the existing data along the isoelectronic sequence (i.e., C I, N II, O III, etc.). In particular, radiative lifetimes in N II and O III were measured with high precision using storage rings [16, 17]. In the present work, we use these exceptionally accurate measurements to identify the most reliable data for both the lifetime of and transition rates from the $2s2p^3\ ^5S°_2$ level of C I.

## 2. Measurements by Goly

Goly [13] determined the *A*-values of several C I transitions by diagnosing the plasma of a wall-stabilized arc in either pure $CO_2$ or $Ar+CO_2$ mixture samples. Relative emission intensities of observed lines were measured on photographic plates. Transition probabilities were derived from these relative intensities under the assumptions that the medium is optically thin and the condition of local thermodynamic equilibrium (LTE) is satisfied. The claimed accuracy of his results was rather low, about 40 % on average. His measured *A*-values for the $2s^22p^2\ ^3P_{1,2} - 2s2p^3\ ^5S°_2$ transitions (lines A and B, respectively) are $(3.5\pm1.6)\times10^2$ s$^{-1}$ and $(1.2\pm0.5)\times10^3$ s$^{-1}$ with a branching ratio B/(A+B) ≈ 0.77. The $^5S°_2$ lifetime derived from Goly's measured *A*-values is $(650\pm210)$ µs.

## 3. Measurements by Ito et al.

Ito et al. [14] measured the radiative rate of the stronger B line by observing the afterglow dynamics of emission from a carbon monoxide plasma generated with a 2.4 GHz pulse-modulated microwave source in an electron cyclotron resonance (ECR) source. Their reported transition rate for this line, $3.9\times10^4$ s$^{-1}$ (with a 30 % uncertainty), was derived from the measured decay rate of the $^5S°_2$ level at various CO pressures in the interval (0.13 to 20) Pa, for which they obtained a mean value of $4.7\times10^4$ s$^{-1}$, corresponding to the lifetime of 21 µs for the $2s2p^3\ ^5S°_2$ level. In the derivation of the transition rate of the B line, Ito et al. used their observed branching ratio B/(A+B) ≈ 0.83. Although Ito et al. did not specify the uncertainty of the measured lifetime, it can be inferred from their discussion to be ≈ 25 %.

## 4. Theoretical data

Prior to measurements of Goly [13] and Ito et al. [14], the most accurate theoretical calculations were those of Garstang [4]. His calculations were made using a non-relativistic Hartree-Fock approximation with superposition of interacting configurations, in which relativistic effects were accounted for perturbatively. The results were 8.0 s$^{-1}$ and 24.0 s$^{-1}$ for the *A*-values of the A and B lines, respectively, with a branching ratio B/(A+B) ≈ 0.75. The $^5S°_2$ lifetime derived from Garstang's *A*-values is 0.031 s.

Mendoza et al. [5] calculated the *A*-values of the $2s^22p^2\ ^3P_{1,2} - 2s2p^3\ ^5S°_2$ transitions (A and B, respectively) for carbon-like atoms and ions with Z = 6–28 in the Breit-Pauli approximation (using the SUPERSTRUCTURE code) and obtained 8.99 s$^{-1}$ and 20.6 s$^{-1}$ for the A and B lines of neutral carbon. The corresponding branching ratio is B/(A+B) ≈ 0.70. The $^5S°_2$ lifetime derived from the *A*-values of Mendoza et al. is 0.034 s.

In 2007, Wiese & Fuhr [11] recommended the theoretical results of Froese Fischer [6] as the best available *A*-values for the A and B lines. Her values, 8.6 s$^{-1}$ and 21 s$^{-1}$, calculated using a multiconfiguration Hartree-Fock approximation with relativistic corrections, agree well with Garstang [4] and with Mendoza et al. [5]. Their accuracy was estimated by Wiese and Fuhr as "D," corresponding to uncertainties < 50 %. The branching ratio of those recommended *A*-values is B/(A+B) ≈ 0.71, and the $^5S°_2$ lifetime is 0.034 s, coinciding with the result of Mendoza et al. [5].

Reviewing the literature, we found an *A*-value for the B line, $5.3\times10^8$ s$^{-1}$, given by Tanaka and Tachibana [8], which was quoted from the compilation of atomic emission lines by Payling and Larkins [15]. This value was obtained by Payling & Larkins from their own nonrelativistic Hartree-Fock calculations. They gave the same *A*-value for both the A and B lines and assigned to both an uncertainty of ±17 %. Thus, the $^5S°_2$ lifetime resulting from calculations of Payling & Larkins is $(0.94\pm0.11)$ ns.

## 5. Discussion

For ease of comparisons, the data on the properties of the A and B lines and the C I 2s2p$^3$ $^5$S°$_2$ level discussed in the preceding sections are collected in Table 1.

Table 1. Data on radiative decay branches and lifetime of C I 2s2p$^3$ $^5$S°$_2$.

| Source | A-value (s$^{-1}$) | | Branching ratio B/(A+B) | $^5$S°$_2$ lifetime $\tau$ (s) |
|---|---|---|---|---|
| | Line A 2964.846 Å Lower level 2s$^2$2p$^2$ $^3$P$_1$ | Line B 2967.224 Å Lower level 2s$^2$2p$^2$ $^3$P$_2$ | | |
| **Experiments:** | | | | |
| Goly [13] | (3.5±1.6)×10$^2$ | (1.2±0.5)×10$^3$ | 0.77 | (6.5±2.1)×10$^{-4}$ |
| Ito et al. [14] | ≈ 8×10$^3$ | (3.9±1.2)×10$^4$ | 0.83 | 2.1×10$^{-5}$ |
| **Theory:** | | | | |
| Garstang [4] | 8.0 | 24.0 | 0.75 | 0.031 |
| Mendoza et al. [5] | 8.99 | 20.6 | 0.70 | 0.034 |
| Froese Fischer [6] (uncertainties from [11]) | 8.6±4.3 | 21±10 | 0.71 | 0.034±0.013 |
| Payling & Larkins [16] | (5.3±0.9)×10$^8$ | (5.3±0.9)×10$^8$ | 0.50 | (9.4±1.1)×10$^{-8}$ |
| **Semiempirical:** | | | | |
| Z-scaling (this work) | – | – | – | 0.043 |

As seen from Table 1, the experimental A-values determined by Goly [13] and by Ito et al. [14] are much greater than the currently recommended data [6, 11] by factors of ≈50 and ≈1400 on average, respectively. The data of Payling & Larkins [15] differ from Refs. [6, 11] by a factor of 4×10$^7$.

To investigate these discrepancies, we made an isoelectronic comparison, for which critically evaluated transition probability data are accessible in the Atomic Spectra Database (ASD) of the National Institute of Standards and Technology (NIST) [18]. We supplemented the ASD data with results of the large scale MCHF/MCDHF (Multiconfiguration Hartree–Fock and Dirac–Hartree–Fock) calculations [19] for carbon-like F, Ne, Mg, and P. Further below, when we mention "reference data," we mean this extended reference data set. Agreement of these reference data with experimental lifetimes was confirmed for N II and O III [16, 17]. The latter measurements were made by observing decay curves of emission from ions circulating in a storage ring. This method has some caveats, but all possible sources of errors were analyzed in detail by Träbert et al. [16, 17], and their results can safely be adopted as a benchmark for checking theoretical data.

Träbert [20] showed that the radiative lifetime of a level decaying by intercombination transitions to levels with the same principal quantum number can be approximated by a simple formula

$$\tau = C/(Z - \mu)^7, \qquad (1)$$

where $C$ is a proportionality constant, $Z$ is the nuclear charge and $\mu$ is a screening constant. For the carbon sequence, fitting of the N II and O III experimental data [16, 17] yields $C = 103(8)$ and $\mu = 2.96(5)$, where the uncertainties given in parentheses are determined by the lifetime measurement uncertainties in N II and O III.

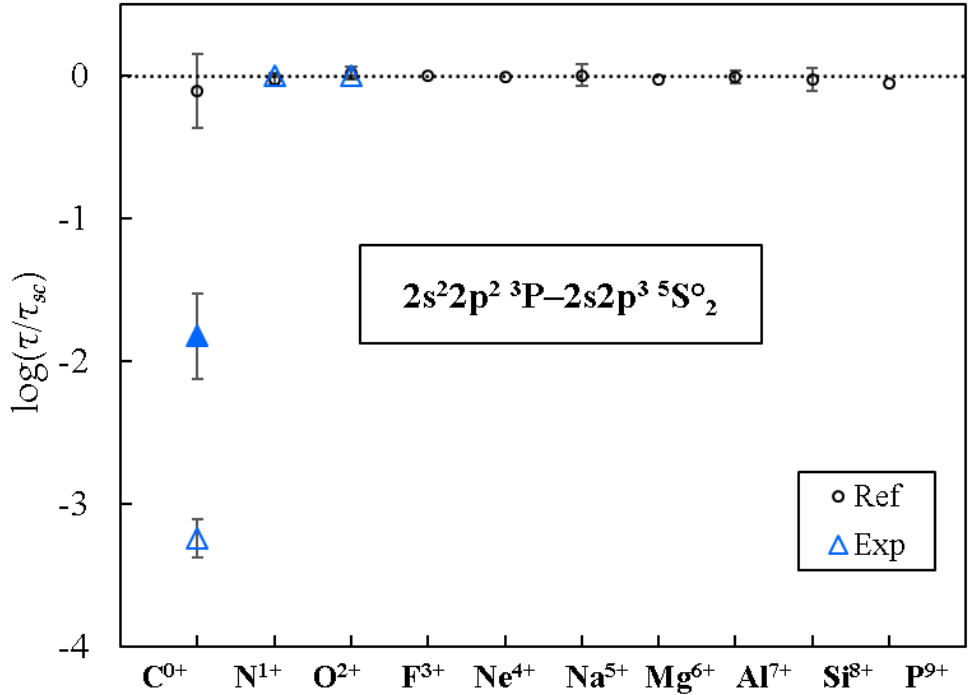

**Figure 2** (*Color online*). Ratios of various lifetime data ($\tau$) to those calculated with Eq. (1) ($\tau_{sc}$) for the $2s2p^3$ $^5S°_2$ level in the carbon isoelectronic sequence. Circles denote the reference data (see text), empty triangles are experimental data for C I, N II and O III [14, 16, 17], and the full triangle is the C I datum from Goly [13]. The error bars represent the measurement uncertainties from the above references and uncertainties of the NIST reference data [18]. The data of Payling & Larkins [15] for C I do not fit on the figure scale and are far below the horizontal axis.

Ratios of radiative lifetimes calculated from the reference data on transition probabilities to those computed with Eq. (1) are plotted in figure 2. The error bars on that figure represent the uncertainties of experimental [13, 14, 16, 17] and NIST [18] data. For C I, the NIST database [18] quotes the theoretical data of Froese Fischer [6]. There are no error bars on the reference data points for F, Ne, Mg, and P [19], since these data have not been evaluated yet. The figure shows that all the reference data agree with Eq. (1) within 11 % on average, which is smaller than the uncertainties of the NIST data [18]. A slightly larger deviation (27 %) is noticeable in C I, but it is still small enough to conclude that the reference data essentially agree with the isoelectronic scaling, which strongly supports their validity. From this comparison, we infer that the theoretical data on F, Ne, Mg, and P [19] have uncertainties smaller than 15 %.

The results of both measurements in C I [13, 14] strongly deviate from both the theoretical reference data and Eq. (1), which suggests that there were some unaccounted errors in those measurements. We investigated both experiments, trying to identify any possible flaws.

We compared all measurements reported by Goly [13] with more accurate determinations involving a normalization of measured branching fractions with respect to observed lifetimes [21]. The comparison indicates that his uncertainty estimates should be at least doubled. Inaccuracy of intensity measurements on photographic plates probably contributed to his measurement uncertainties, but it still cannot explain the difference from reference data by a factor of ≈50. A possible explanation is that the LTE condition, essential for his derivation of *A*-values, was not satisfied for the $^5S°_2$ level. As Goly derived from his measurements, the plasma temperature, electron density, and ambient gas density were $T \approx 11000$ K, $n_e \approx 5\times10^{16}$ cm$^{-3}$, and $n \approx 10^{17}$ cm$^{-3}$, supporting his LTE assumption. Perhaps, the lines from the $^5$S°$_2$ level originated from outer regions of the plasma column having lower electron density, allowing this level to accumulate an excess population due to its metastability. However, there are no measurements of spatial distributions of the plasma parameters to support this explanation. Regarding the relative intensities of the A and B lines, we note that Goly's observed intensity ratio, B/(A+B) ≈ 0.77, is not affected by the population model. These lines were observed with a similar intensity ratio by Shenstone [1] and Johansson [2] (0.71 and 0.67, respectively).

We examined possible systematic effects in the experiment of Ito et al. [14], but could not find any plausible explanation for the discrepancy between their results and theory [4–6, 11].

The discrepant theoretical result of Payling & Larkins [15] is also difficult to explain. To understand the general quality of their data, we compared all their C I *A*-values with recommended values [18]. The compilation [15] contains 1148 atomic C I emission lines in the range of (101.7 to 897.7) nm derived from known energy levels, with transition probabilities accompanied by accuracy estimates and intensities. Even excluding a few possible misprints, we found the transition wavelengths given by Payling & Larkins to be very inaccurate. For the 960 C I lines, the standard deviation from our Ritz values [12] is 0.6 Å with a few as large as 4.0 Å, and many *A*-values deviate from the NIST recommendations [18] by several orders of magnitude. The *A*-value accuracy estimates of Payling & Larkins are labeled as "m" = (10 to 40) % and "a, b, c, d, e" = 2 %, 3 %, 6 %, 11 %, and 17 %, respectively. The *A*-values given with the label "m" were taken from measurements reported in an early version of the NIST ASD [18] released in 1998 and from other sources dating before 1990. The other labels correspond to calculations made by Payling & Larkins. They estimated their uncertainties by comparing the calculated *A*-values with measured ones. We re-evaluated them by comparing the line strengths $S_{PL}$ computed from the *A*-values of Payling and Larkins with the corresponding critically evaluated $S_{ASD}$ data from the NIST compilations [10, 11]. The technique involved in this process is the same as detailed in several other works [22–25]. A total of 325 matching transitions were used in this comparison. Such comparisons were made separately for transitions of different type (for example, intercombination transitions and/or those involving the change of the orbital angular momentum $\Delta\ell > 1$). As this analysis showed, the mean uncertainty is 20 % for *A*-values with accuracy labeled as "m," while for other transitions preserving spin the uncertainties greatly depend on the magnitude of line strength $S_{PL}$. For weak transitions with $S_{PL} < 0.1$, uncertainties are as large as two or three orders of magnitude. For intercombination transitions, the *A*-values of Payling & Larkins deviate from the NIST recommendations by a few orders of magnitude on average, regardless of their strength. In short, consistency of C I data given by Payling & Larkins is poor, and hence they are not of much use.

## 6. Conclusion

The previously reported measurements of the transition rates from the 2s2p$^3$ $^5$S°$_2$ level of neutral carbon were critically analyzed and compared with theoretical results. Inconsistencies in the observed values are tested against isoelectronic data. This comparison strongly supports the currently recommended theoretical data [6, 11] and indicates that both observations [13, 14] in C I were distorted by some systematic effects. No convincing explanations were found for the discrepant measurements of Goly [13] and Ito et al. [14]. As a side result, we validated the radiative lifetimes of the 2s2p$^3$ $^5$S°$_2$ level calculated by Froese Fischer and Tachiev [19] for C-like F, Ne, Mg, and P and estimated their accuracy to be better than 15 %. Spectroscopic data on C I given in the compilation by Payling & Larkins [15] was found to be of poor quality. To resolve the existing discrepancies, a new experimental measurement of radiative lifetime of the 2s2p$^3$ $^5$S°$_2$ level of neutral carbon is highly desirable.

**Acknowledgements**
This work is partially funded by the Astrophysics Research and Analysis program of the National Aeronautics and Space Administration of the USA. Authors gratefully acknowledge valuable discussions with Yuri Ralchenko, NIST. K. Haris was working at NIST under a Guest Researcher agreement 131227.